\documentclass[a4paper,14pt]{extarticle}
\usepackage{graphicx}
\usepackage{color}
\usepackage{amsmath}
\usepackage{amssymb}
\usepackage{wrapfig}
\usepackage{hyperref}
\usepackage{soul}
\usepackage{epstopdf}
\usepackage{amsfonts}
\usepackage{amsmath}

\begin{document}

\title{Big fish and small ponds: why the departmental h-index should not be used to rank universities \thanks{This work was supported in part by the National Academy of Sciences of
Ukraine, Project KPKBK 6541230.}}

\author{O.~Mryglod$^{1,2}$, Yu.~Holovatch$^{1,2}$, R.~Kenna$^{2,3}$}

\date{%
$^1$	Institute for Condensed Matter Physics of the National Academy of Sciences of Ukraine, 1 Svientsitskii St., 79011 Lviv, Ukraine; \\
	$^2$ ${\mathbb L}^4$ Collaboration \& Doctoral
College for the Statistical Physics of Complex Systems, Leipzig-Lorraine-Lviv-Coventry, Europe\\
$^{3}$
Centre for Fluid and Complex Systems, Coventry University, Coventry, CV1 5FB, United
Kingdom
	\today
}


	\maketitle

\begin{abstract}
	The size-dependent nature of the so-called group or departmental \emph{h-index} is reconsidered in this paper.
	While the influence of \textcolor{black}{unit} size on such collective measures  was already demonstrated a decade ago,  institutional ratings based on
this metric can still be found and still impact on the reputations and funding of many research institutions.
	The aim of this paper is to demonstrate the fallacy of this approach to collective research-quality assessment in a simple way, focusing on the \emph{h-index} in its original
form.
	We show that randomly reshuffling real scientometric data (varying numbers of citations) amongst institutions of varying size, while maintaining the volume of their research
outputs, has little effect on their  departmental \emph{h-index}.
	This suggests that the relative position in ratings based on the collective \emph{h-index} is determined not {only}  by quality (impact) of particular research
outputs but by their volume.
	Therefore, the {application of the} collective \emph{h-index} in its original form is {disputable} as a basement for comparison at aggregated
levels such as to research groups, institutions or journals.
	We suggest a  possible remedy for this failing which is implementable in a manner that is as simple and understandable as the \emph{h-index} itself.

\end{abstract}

%

\section{Introduction}\label{sec0}
\emph{Camille No$\hat{u}$s} is a fictitious character who emerged into the academic world in 2020 {\cite{Camille}} (see also \cite{krasnytska2020ising}).
She was conceived as a ``collective individual'' to embody research of the community as a whole, representing a desire to return to traditional approaches to construction and
dissemination of knowledge  ``under the control of the academic community''.
Her forename is an amalgam of the collegial ``we'' (``nous'' in French) and the concepts of reason, mind or intellect (``$\nu$o$\tilde{\nu}$$\zeta$'')
from Greek philosophy.
Symbolizing the ``deep attachment'' to traditional academic values, \emph{Camille} is the antithesis of ``indicators elaborated by the institutional management of
research''.
Standing for the collective and not the individual, she is in contrast to the increased use of  metrics in the management and control of research.
Indeed, this contrast between the collective and the individual is one that is frequently lost
{ in an increasingly corporatised academic world.
The increasing variety of science-related activities beyond research itself causes the engagement of more and more professional managers, university leaders
and other professionals who do not have personal academic experience.
Sometimes  overly simplified understanding of the underlying processes, together with a lack of  knowledge of statistics and the need to seek higher positions
 in league tables in a highly competitive academic world, may lead to oversimplified interpretation and, thus, misuse of scientometric indicators and approaches.
	Tools based on statistics that are designed to measure the academic performance of the collective are applied to develop targets for the individual.
	And tools designed to measure one element of an individual academic's achievements are adapted for ranking and therefore funding the collective.
	Altogether, this drives human pursuits of knowledge in a manner very different to the traditions that Camille represents.}

Here we revisit one such tool -- the departmental \emph{h-index} calculated in a straightforward manner.
The \emph{h-index} itself is probably the most popular and, at the same time, the most debated scientometric indicator \cite{hirsch2005index,alonso2009h}.
More than 85 variants of it have been developed and many are listed in  \cite{schubert2019all}.
Besides numerous modifications of the index itself,
the generalization of this indicator of {individual} researchers' performance to {{various}} levels of aggregation such as groups, institutions,
countries, journals or scientific topics was suggested from the very beginning.
The most straightforward definition of {the} group \emph{h-index} is 
the same as the original one, but {involves} all publications by group members in {a single} pool.
{The}  group \emph{h-indices} discussed further in this paper are calculated in this way.
``Crude and imperfect'' {as it is, it is}  sometimes suggested as an alternative {to other research evaluation frameworks, e.g., the UK's
REF\footnote{{The {\emph{Research Excellence Framework}} (REF, seehttp://www.ref.ac.uk) is a peer-review based regular assessment of the quality of research emanating
from universities and higher education institutes (HEI) in the UK.}}}, as a means to  rank departments in at least some disciplines.
{A famous example is that of influential Oxford academic Dorothy Bishop in Ref.~}\cite{Dorothy}.
Other examples include ratings of Ukrainian Higher Educaton and Research Institutions.
{These are fully} based on group \emph{h-index} {and} are periodically updated,  {see Refs.~}\cite{BibliometrykaUkr,OsvitaUkr}.
Although Bishop's suggestion \cite{Dorothy} was countered in the famous Metrics Tide report which went on to help shape the UK's research landscape, and although these scores are
not officially implemented into governmetal assessment procedures in Ukraine, they remain persuasive amongst research managers and some funding bodies.
So, even {though they are} groundless, they can be eventually perceived as objective.
{This is a classic example of} the Thomas Theorem {in sociology} \cite{bornmann2020thomas}.

We used very simple methods to show that aggregate measures of \emph{h-indices} are not suitable indicators of the quality or research emanating from research groups, departments,
or indeed nations.
{To pitch this at the level of non-experts, a very simple statistical process  was used and explained}.
We showed that the collective (departmental or group) h-index, as used widely is {to a large extent}  an indicator of the \emph{number of outputs generated by a given
institution, a value that itself is a proxy for size}.
This may be ascribed to the simple but mischievous statistical phenomenon of sampling bias;
to use an analogy from angling, if more big fish are desired, use a big net to catch them;
if randomly distributed, there are fewer big fish in small ponds than there are in big ones.
Random sampling from a randomly shuffled pool of research outputs makes it more likely that highly cited papers are ascribed to larger universities.
If the  \emph{h-index} is to be used in any meaningful way as a collective metric, it therefore should be normalised according to size.


Table~\ref{tabl1a} illustrates the persuasive power of the departmental \emph{h-index}.
It lists the top and bottom UK universities ranked according to their  departmental \emph{h-index} scores in physics as extracted using Scopus data
(the full table is presented as Table~\ref{tabl2} below.)
As might be expected, the most prestigious universities such as Oxford and Cambridge are ranked highest and lesser known universities are ranked lowest.
The fact that this table captures at least a qualitative correlation between prestige and the departmental \emph{h-index} suggests the latter as a reasonable tool for measuring
research quality.
It reinforces the notion that the \emph{h-index} is a reasonable way to rank institutions.
\begin{table}[ht]
	\caption{Number of publications and values of group \emph{h-indices} for Higher Education Institutions (HEI) in
		UK (Subject Area: ``Physics'', publication window: 2001--2007, accessed in March 2020).
}
	\begin{center}
		{\small
			\begin{tabular}{|p{5cm}|p{5cm}|}			
				\hline
				HEI & Scopus group \emph{h-index}\\
				\hline \hline
				University of Cambridge&250\\
				University of Oxford&207\\
				Imperial College London&203\\
				$\dots$ & $\dots$\\
				Aberystwyth University&40\\
				University of the West of Scotland&31\\
				University of Brighton&29\\
				\hline
			\end{tabular} \label{tabl1a}
		}
	\end{center}
\end{table}

The perceived link between the departmental \emph{h-index} and prestige might be seen to suggest that the attractiveness and reputations of prestigious institutions are the very
things that draws the highest quality to them; the more attractive a university is,  the better quality people they draw and the higher quality the research is.
It is undoubtedly true that quality people are drawn to quality institutions but is this really what is being captured by the departmental \emph{h-index}?
If these research outputs were to be re-shuffled and randomly assigned to various institutions, one might expect a different picture to emerge --- a more egalitarian one.
I.e., a person ill-versed in the notion of bias in statistics might expect that random allocation of all research outputs of a nation to its various universities would lead to
random rankings of these universities.
We did precisely this --- we randomly reshuffled research outputs amongst research institutions and ranked them according to their resulting departmental \emph{h-indices}.
{It might be surprising that the resulting scores were not so ``random'' as it could be expected.}

The remainder of this paper is devoted to explaining this phenomenon as due to size bias.
It is organized as follows.
We provide some context to the extensive nature of the \emph{h-index} in the next section.
In Section ``Combinatorics of scale'' we demonstrate the principal danger of usage of absolute values dealing with statistical predictions for the number of balls in baskets of
varying sizes while \emph{h-index} is based on absolute values by its definiton. In the Section titled ``Shaking out the correlations'' we calculated the expected group
\emph{h-index} for research units (groups) of different sizes based on real but reshuffled citation data. This experiment is performed for two case studies, discussed in two
subsections, correspondingly. The resulting shapes of characteristic curves are discussed. A short summary is presented in the last section.

\section{Context: Size matters in scientometrics}\label{sec1}

The dependency of the \emph{h-index} {at the level of units which size is identified by the number of publications} was
discussed already over decade ago \cite{glanzel2006h,egghe2006informetric} in the context of journals \cite{schubert2007systematic}, institutions
\cite{molinari2008new,sypsa2009assessing}, disciplines \cite{kinney2007national} or countries \cite{babic2016evaluation,montazerian2019new}.
In each case, the growth of \emph{h-index} with  number of publications $N$ was found to be:
\begin{equation}
h \sim N^{\beta}, \qquad  N\gg 1.
\label{eq1_beta}
\end{equation}
The quantity $\beta>0$ was found to be similar for different levels of aggregations (different {units} such as journals, institutions, etc.)
\cite{molinari2008new,kinney2007national,sypsa2009assessing,schubert2007systematic}.
The reasons for such universality have been widely discussed, with the presumption that the distribution of citations falls in the family of Paretian distributions with power-law
exponent
$\alpha$, leading to an estimate $\beta \approx \frac{1}{1+\alpha}$, see \cite{glanzel2006h,glanzel2019springer}.
It is interesting to note, that such universalities were found independently of the manner in which  the input subsets were built:
\begin{itemize}
	\item by assembling publications published in different years (cumulative process)
	\cite{molinari2008new,kinney2007national};
	\item by considering particular subsets of general collections: e.g., journal publications  related to different countries or publications assigned to different subject
categories  \cite{molinari2008new};
	\item by considering independent research units of disparate sizes \cite{sypsa2009assessing,ye2009investigation}.
\end{itemize}

This paper fits into the above body of work.
Its goal is to illustrate the {significant} role that research-unit size plays in the collective \emph{h-index}.
%
The number of published outputs can be considered as a proxy for the size of research unit
due to natural dependency of the number of outputs on the number of active researchers (to give an example, Pearson correlation coefficient for the number of documents in Scopus vs
corresponding number of authors equals to 0.98 for the first case study considered in this paper, i.e., for Ukrainian HEIs) \footnote{Of course, the number of publications can serve
only as a proxy for the size of research unit since many factors such as a number of grants or level of basic financial support can have an impact.}.
This illustration adds to a growing body  of evidence that the \emph{h-index} is not a good basis for {straightforward} comparison or ranking research groups,
institutions or countries.
{This doesn't mean that \emph{h-index} calculated for any collection of documents loses {meaning}.
However, the results have to be interpreted in a proper way.
Departmental \emph{h-index} {is not solely a function of}  quality of outputs and, therefore, the rating based on such score is not a
quality rating.}

By basing our illustration on a very simple combinatorics model, we hope to {make it}  accessible to a range of scientific managers.
Here we solely focus on the statistical properties the group  \emph{h-index};
other arguments such as the innate inconsistency of the group \emph{h-index} itself {as addressed}  in  \cite{waltman2012inconsistency} are not discussed here.

%
{Our findings lend support to the idea that the assessment of performance of research groups has to be relative, i.e., made rather in respect to some benchmark. Similiarly as relations  between group h-indices and the numbers of contributions by different countries are built for several journals in \cite{molinari2008new}, the corresponding relations can be derived for any collection of cited publications. Besides the empirically observed relations, their version for the uncorrelated system can be observed \cite{crespo2012citation,Meyers2017}.}
%
%
%

\section{Combinatorics of scale}\label{sec2}

The aim of this section is to explain the phenomenon described above in the simplest possible manner.
To this end we neglect all the emergent effects discussed elsewhere in the literature, such as  peculiarities of local managements, infrastructures,  recruitment strategies, economy
of scale, the Matthew effect and so on. {We rather attract attention to the role of the {sample} size 
	suggesting a simple intuitive way to separate this factor and assess its contribution.}
We imagined having a data set to hand which comprises of a collection of research outputs {(publications)} and their citation counts for each individual unit to be
assessed (i.e., for each individual research group or university).
Each unit produced a certain amount of outputs and we termed the number of {published} outputs by a given institution  its ``productivity''.
Productivity contains no information about quality or citation counts of individual papers.
We imagined placing all of these outputs (each of which has their own citation counts) into  single ``pot'', ``pool'' or ``basket''.
We then ``{fish}'' them out again and {distribute} them randomly amongst the various institutions, maintaining the institutions original productivity.
In this way citation counts were randomly distributed.
The only difference between institutions for this purpose was the size of their individual ``basket''.
Then we measured the collective or departmental \emph{h-index} of each institution.
Our {challenge then was} to determine the effect of this redistribution of wealth on the departmental \emph{h-indices}.
If size plays no role in the \emph{h-index}, one would expect rankings of institutions to be randomly changed.
The extent to which this does not happen provides a measure of the extent to which size matters for the \emph{h-index}.
We will see {in the next sections} that by maintaining their ``basket'' size institutions largely maintain their position in the rankings so that size indeed plays a
predominant role.
In other words, {our results show that} the \emph{h-index} is not a {straightforward} way to compare institutions of
different sizes.
To provide a simple mathematical basis to understand {this phenomenon, we studied} the selection of a given number of balls of different sorts from a pool of balls.
This is a typical task of combinatorics.
We considered $k_1$ black and $k_2$ white balls in a pool containing $K=K_1+K_2$ balls.
The number of combinations that contain exactly $k_1$ black and $k_2$ white balls is
\begin{equation}
	{ \cal N}_{K_1,K_2}^{k_1,k_2}=\mathrm{C}_{K_1}^{k_1}\cdot \mathrm{C}_{K_2}^{k_2}, \qquad
	{\text{where}} \qquad \mathrm{C}_{K}^{k}=\frac{K!}{k!(K-k)!}\,
\end{equation}
and where $k_1+k_2=k$ is the size of the basket.
Therefore, the probability of fishing out $k_1$ black and $k_2$ white balls from a pool of
$K$ balls using a basket of size $k$ is
\begin{equation}
P_{K_1,K_2}^{k_1,k_2}=\frac{{C}_{K_1}^{k_1}\cdot {C}_{K_2}^{k_2}}{\mathrm{C}_{K}^{k}},
\end{equation}
where $\mathrm{C}_{K}^{k}$ is the number of all possible combinations independent of ball color.

\begin{figure}[h]
\centerline{\includegraphics[width=0.8\textwidth]{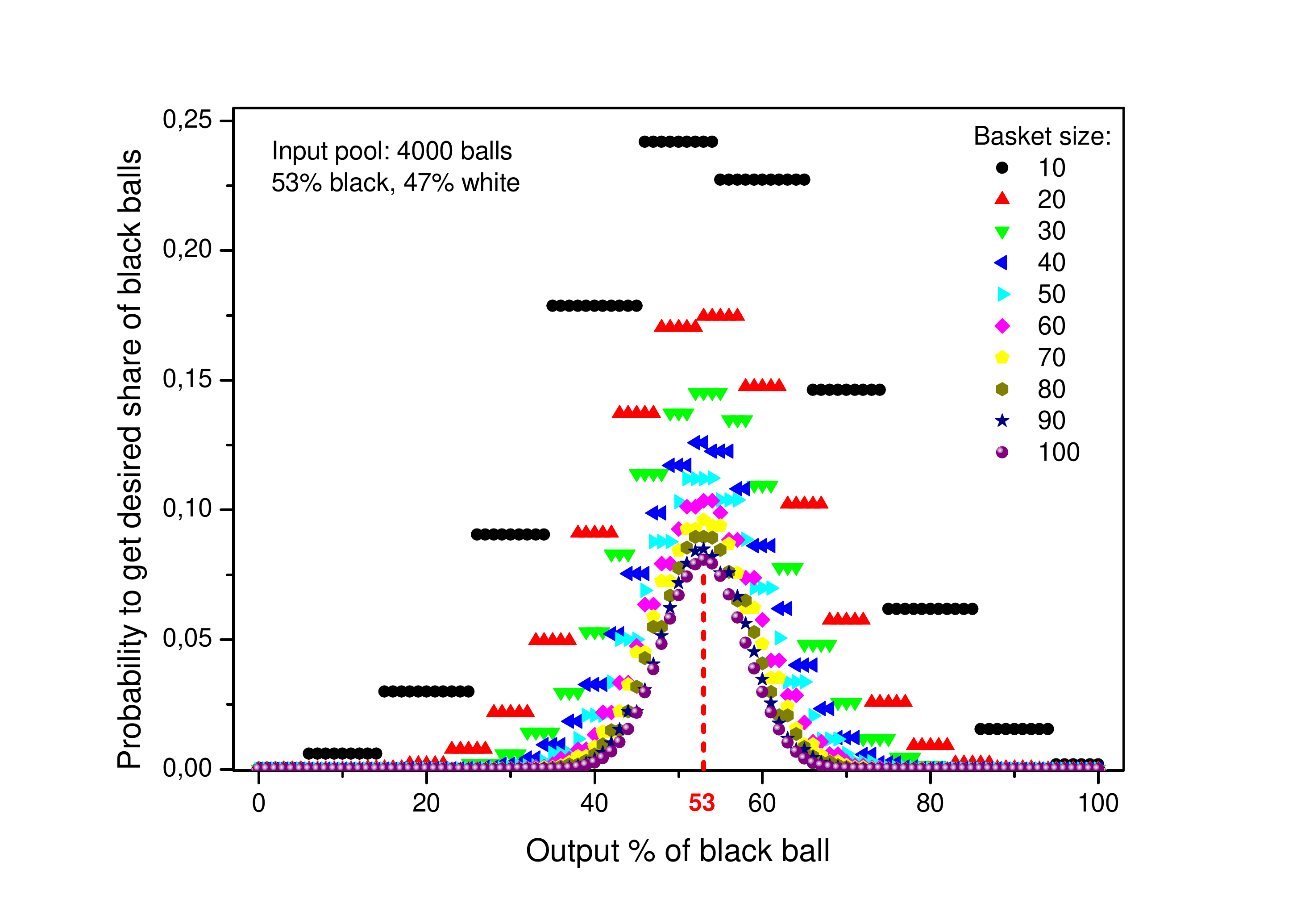}}%
	\caption{Probability to get a given share ($k_1/k$) of black balls  from the pool with the same share ($k_1/k=K_1/K$) using a basket of size $k$. In this figure: $K=4000$,
$k_1/k=K_1/K=53\% $, $k=10, 20, \dots, 100$.}
\label{Fig_perc_black}%
\end{figure}

Obviously the proportion of black and white balls picked up by a basket of size $k$ {is} $K_1/K_2$ on average  (except the cases when basket size is too small to reach the most
probable proportion with integer number of balls).
For example, if 53\% of balls in the pool were black, the most probable proportion of black balls in the basket {is} also 53\%.
Fig.~\ref{Fig_perc_black} illustrates this for baskets of different sizes and a pool of 4000 balls.
Clearly, there is a distribution of ratios around the average value of 53\% and that distribution is less ``granular'' as the basket increases in size.

\begin{figure}[h]
\centerline{	\includegraphics[width=0.8\textwidth]{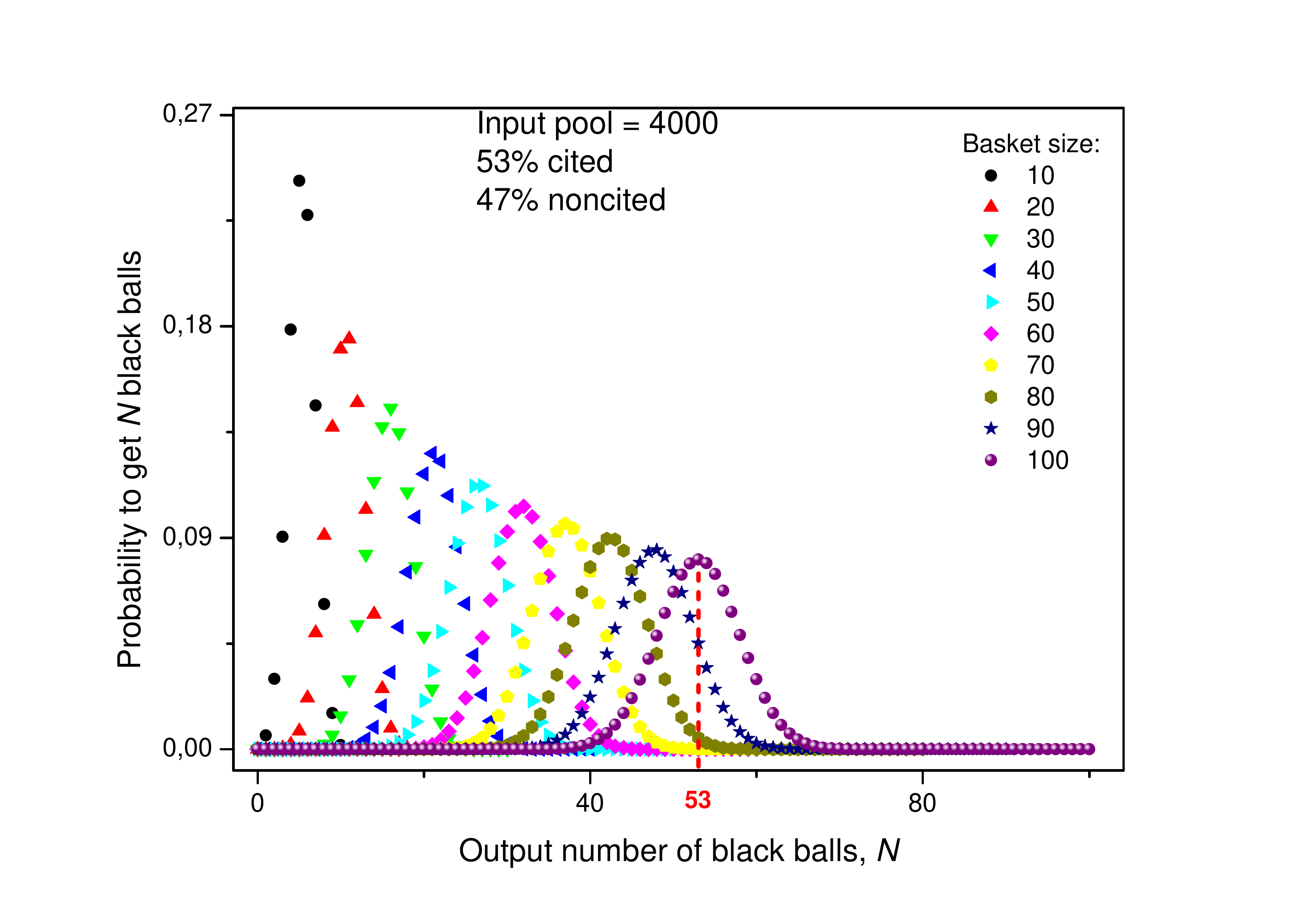}}%
		\caption{Probability to get $k_1$ black balls from the pool with the black balls share $K_1/K$
			using a basket of size $k$. In this figure: $K =4000$, $K_1/K=53\% $, $k=10, 20, \dots, 100$.}
		\label{Fig_num_black}
\end{figure}

Instead of the proportion of black balls, one can explore their absolute number.
In Fig.~\ref{Fig_num_black} we plotted the probability to fish out $N$ black balls from the same pool of 4000 balls (53\% of which are black) using baskets of different sizes.
Obviously a basket of size 10 cannot fish out more than 10 black balls and, as $N$ increases the number of balls fished out increases.
becomes obvious, that larger baskets get proportionally more black balls, see

We considered institutions of different sizes as the baskets of different capacity in the experiments discussed in the next sections.
Thus, we had `baskets' (institutions) full of `balls' (papers) of different `values' (citations). Putting `balls' from all `baskets' together, reshuffling and redistributing them
over the `baskets` again we {lose} any correlations between publications and any additional factors remaining solely with the sizes. 
To check what would happen, two
case studies have been performed, see the next subsections.

\section{Shaking out the correlations. Case studies: {h}-index rankings in Ukraine and the UK}\label{sec3}


\subsection{Case study 1: 40 HEIs in Ukraine}\label{sec4UKR}

Lists of research Institutions in Ukraine, sorted by group \emph{h-indices},  are periodically published by different sources using titles such as as ``Rating of Universities and
Research Institutions'' \cite{BibliometrykaUkr,OsvitaUkr}.
We gathered bibliographic and citation data for 40 such Higher Education Institutions (HEI's) from the Scopus database for  May 2019.
We {do} not fix citation or publication windows {and}
we do not distinguish { different types of} paper {or} citations
since these {details} are not used for the rating scheme either.
Likewise, we used publications from to the full Scopus profile of Institute.
The data for the 40 institutions are listed in Table~\ref{tabl1}.
\begin{table}[ht]
	\caption{Number of publications for 40 Ukrainian HEIs and the corresponding group \emph{h-index} values based on  Scopus data {(}accessed in May 2019{).} {For data
source,} see \cite{UkrData}.}
	\begin{center}
		{\tiny
			\begin{tabular}{|p{9cm}|p{1cm}|p{1cm}|}			
				\hline
				HEI & Number of publications & Group \emph{h-index}\\
				\hline \hline
				Taras Shevchenko National University of Kyiv&17349&90\\
				V. N. Karazin Kharkiv National University&9452&70\\
				Ivan Franko National University of L'viv&6660&61\\
				Odessa I.I.Mechnikov National University&3469&61\\
				Yuriy Fedkovych Chernivtsi National University&3437&61\\
				National Technical University of Ukraine ``Igor Sikorsky Kyiv Polytechnic Institute''&7748&54\\
				Donetsk State Medical University&1255&46\\
				National Technical University Kharkiv Polytechnic Institute&3775&43\\
				Oles Honchar Dnipro National University&3671&43\\
				Danylo Halytsky Lviv National Medical University&1025&42\\
				Lviv Polytechnic National University&6338&42\\
				Sumy State University&2339&39\\
				Vasyl Stefanyk Precarpathian National University&688&38\\
				Uzhgorod National University&2245&37\\
				Ukrainian State Chemical Technology University&1120&36\\
				State Establishment Dnipropetrovsk Medical Academy of Health Ministry of Ukraine&275&35\\
				The Bohdan Khmelnytsky National University of Cherkasy&455&35\\
				V.I. Vernadsky Crimean Federal University&2622&34\\
				Bogomolets National Medical University&741&33\\
				National University of Kyiv-Mohyla Academy&512&32\\
				Vasyl' Stus Donetsk National University&1817&32\\
				National Aerospace University ``Kharkiv Aviation Institute''&1303&30\\
				Kharkiv National University of Radio Electronics&3030&29\\
				Lesya Ukrainka Eastern European National University&771&29\\
				Kharkiv National Medical University&596&27\\
				National University of Life and Environmental Sciences of Ukraine&872&26\\
				Donetsk National Technical University&1336&25\\
				Sevastopol State University&1128&24\\
				National Aviation University&2033&21\\
				National University of Food Technologies of Ukraine&559&21\\
				Kyiv National University of Technologies and Design&486&20\\
				Odessa National Polytechnic University&915&20\\
				K. Ushynsky South Ukrainian Pedagogical University&306&19\\
				Donbass State Engineering Academy&377&18\\
				Ukrainian National Forestry University&240&18\\
				Khmelnytsky National University&419&17\\
				Kryvyi Rih National University&429&17\\
				Odessa State Medical University&387&15\\
				Pridneprovskaya State Academy of Building and Architecture&150&14\\
				Volodymyr Dahl East Ukrainian National University&503&13\\
				\hline
			\end{tabular} \label{tabl1}
		}
	\end{center}
\end{table}

\begin{figure}[ht]
	\centerline{\includegraphics[width=0.8\textwidth]{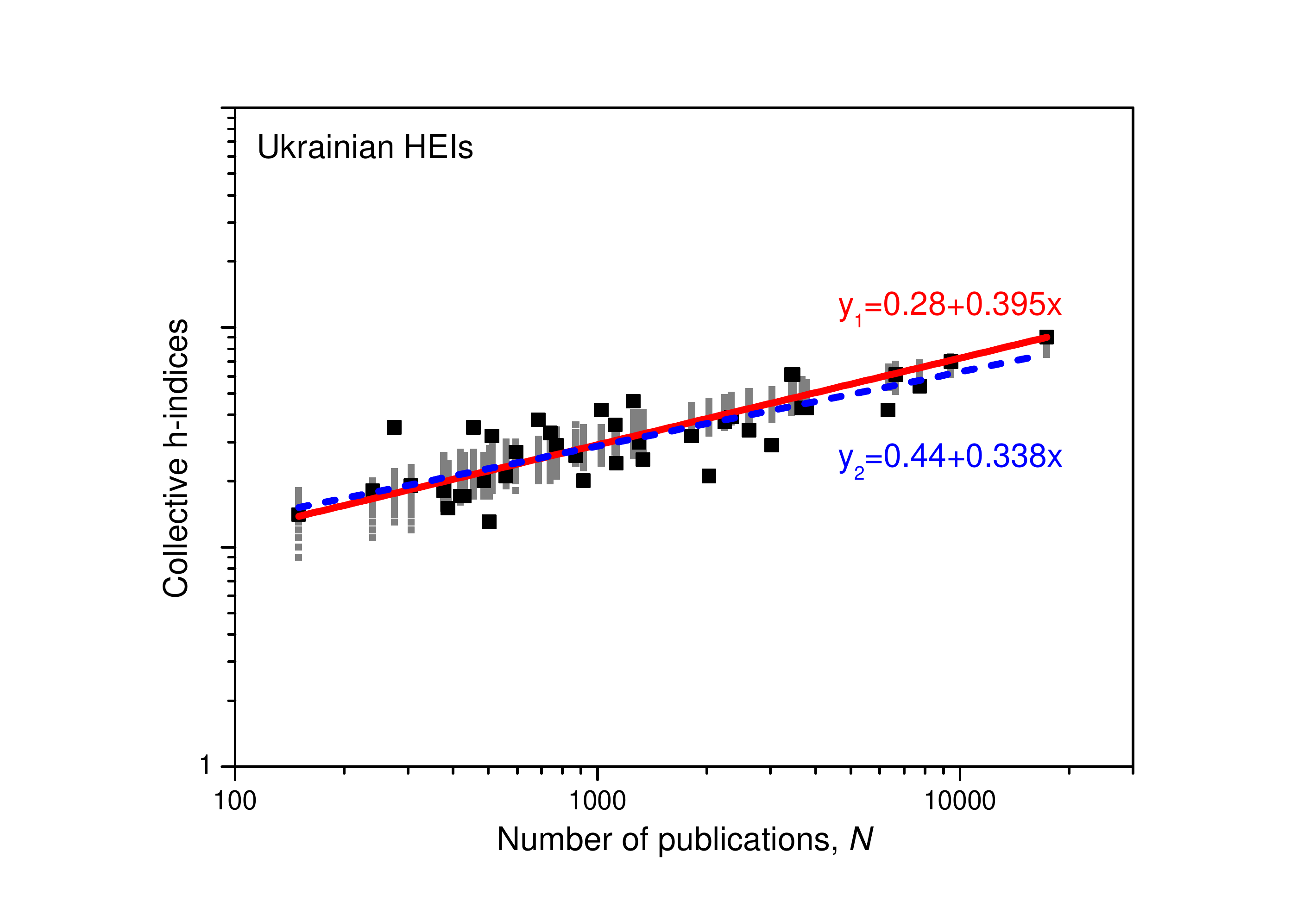}}
	\caption{Values of group \emph{h-indices} {plotted against} group sizes for 40 Ukrainian HEIs.
 {Reshuffled data  are presented in grey}  and {real data are in black}. Both data {sets} tend to follow a power-law dependency, Eq.~(\ref{eq1_beta}).
 {The solid red line {is a fit to}  all reshuffled data put in one bag implying $\beta \approx 0.395$ while
 the broken blue line is for real data with $\beta\approx 0.338$. Both results are statistically significant for $\alpha=0.01$}.}
	\label{Fig_hreshuffled_MON}
\end{figure}

We randomly re-allocated all outputs across the 40  institutions ({over} 90 thousand papers\footnote{including repetitions for the cases of papers co-authored by researchers
from different institutions}), keeping their productivities fixed.
The reshuffling was performed as follows:
\begin{itemize}
	\item the number of publications for each HEI $N$ {was} recorded;
	\item all the publications, labelled by their citation counts, {were put in one pool};
	\item all the publications {were} redistributed among HEIs randomly maintaining their original numbers of publications;
	\item {the group \emph{h-index} for each HEI was calculated}.
\end{itemize}
After {200} reshuffligns we had the set of \emph{h-indices} based on randomly distributed data and repeatedly calculated for each HEI.
The outcomes {are plotted in  Fig.~\ref{Fig_hreshuffled_MON}.
There, t}he horizontal axis represents the number of publications generated by each of 40 institutions.
The vertical axis for the black data are the real institional \emph{h-indices}.
The grey data points are the corresponding values after  reshuffling.
One {can} observe a strong correlation between  \emph{h-indices} and the number of publications for real and randomised data.
An average Spearman correlation coefficient of $\approx 0.77$ indicates {significant}  correlation between ratings based on reshuffled data and the original one.

If follows from Fig.~\ref{Fig_hreshuffled_MON} that the reshuffled values can be approximated by a line {on a} log-log scale with the slope close to $\beta\approx 0.4$,
see \eqref{eq1_beta}. {The corresponding value of $\beta$ estimated for real data is similar. This} value was found for a number of real data collections with
$10^2<N<10^5$ \cite{molinari2008new}.

\subsection{Case study 2:  41 research groups in Physics in {the} UK}\label{sec4UK}

{While in the previous} case study  we applied the reshuffling procedure to data from different institutions, irrespective of the field of science they
represent{, in case study 2 we analyse} data {from} the same field of science.
To this end, we used Scopus data for 41 research groups working in the field of Physics and submitted to the UK's {last} Research Assessment Exercise
\footnotemark{}\footnotetext{ {The} \emph{Research Assessment Exercise} (RAE, see http://www.rae.ac.uk/) {was} the predecessor of  {\emph{Research Excellence Framework}}
(REF).}.
 Publications {were} limited to {the years} 2001--2007  (see Table~\ref{tabl2}) {because that was the timeframe to which RAE2008 pertained}.
 {It also leads to results comparable with others published before (e.g., \cite{mryglod2013absolute}).}

\begin{table}[ht]
	\caption{Number of publications and values of group \emph{h-indices} for 41 HEIs in
		UK submitted to RAE2008 {the} UOA (unit of assessment) {categorised as} ``Physics''. The group
		\emph{h-index} values are based on  Scopus data accessed in March 2020, see \cite{UKData}.}
	\begin{center}
		{\tiny
			\begin{tabular}{|p{7cm}|p{3cm}|p{1cm}|}			
				\hline
				HEI & Number of publications [2001-2007]; 'Physics'& Scopus \emph{h-index}\\
				\hline \hline
				University of Cambridge&9602&250\\
				University of Oxford&8129&207\\
				Imperial College London&7186&203\\
				University of Durham&3208&178\\
				University College London&4842&165\\
				University of Edinburgh&2947&156\\
				University of Manchester&5588&156\\
				University of Southampton&4274&150\\
				University of Bristol&3104&149\\
				University of Sheffield&3389&136\\
				University of Birmingham&2918&133\\
				University of St Andrews&2003&130\\
				University of Nottingham&2609&128\\
				University of Leeds&2687&126\\
				University of Glasgow&2604&125\\
				Queen Mary, University of London&2090&124\\
				University of Liverpool&2966&123\\
				University of Sussex&1398&117\\
				University of Leicester&1571&114\\
				Queen's University Belfast&1932&108\\
				The University of Warwick&2226&99\\
				Cardiff University&1382&98\\
				University of Bath&1268&92\\
				University of Exeter&1411&92\\
				University of Surrey&2360&92\\
				Liverpool John Moores University&616&91\\
				University of Strathclyde&2115&90\\
				Lancaster University&1243&87\\
				Loughborough University&1455&83\\
				Heriot-Watt University&1358&79\\
				Royal Holloway University of London&664&78\\
				University of Hertfordshire&680&78\\
				King's College London&1117&77\\
				University of York&1211&69\\
				Swansea University&753&68\\
				Keele University&585&63\\
				University of Kent&578&61\\
				University of Central Lancashire&283&47\\
				Aberystwyth University&250&40\\
				University of the West of Scotland&176&31\\
				University of Brighton&117&29\\
				\hline
			\end{tabular} \label{tabl2}
		}
	\end{center}
\end{table}

\begin{figure}[ht]
	\centerline{\includegraphics[width=0.8\textwidth]{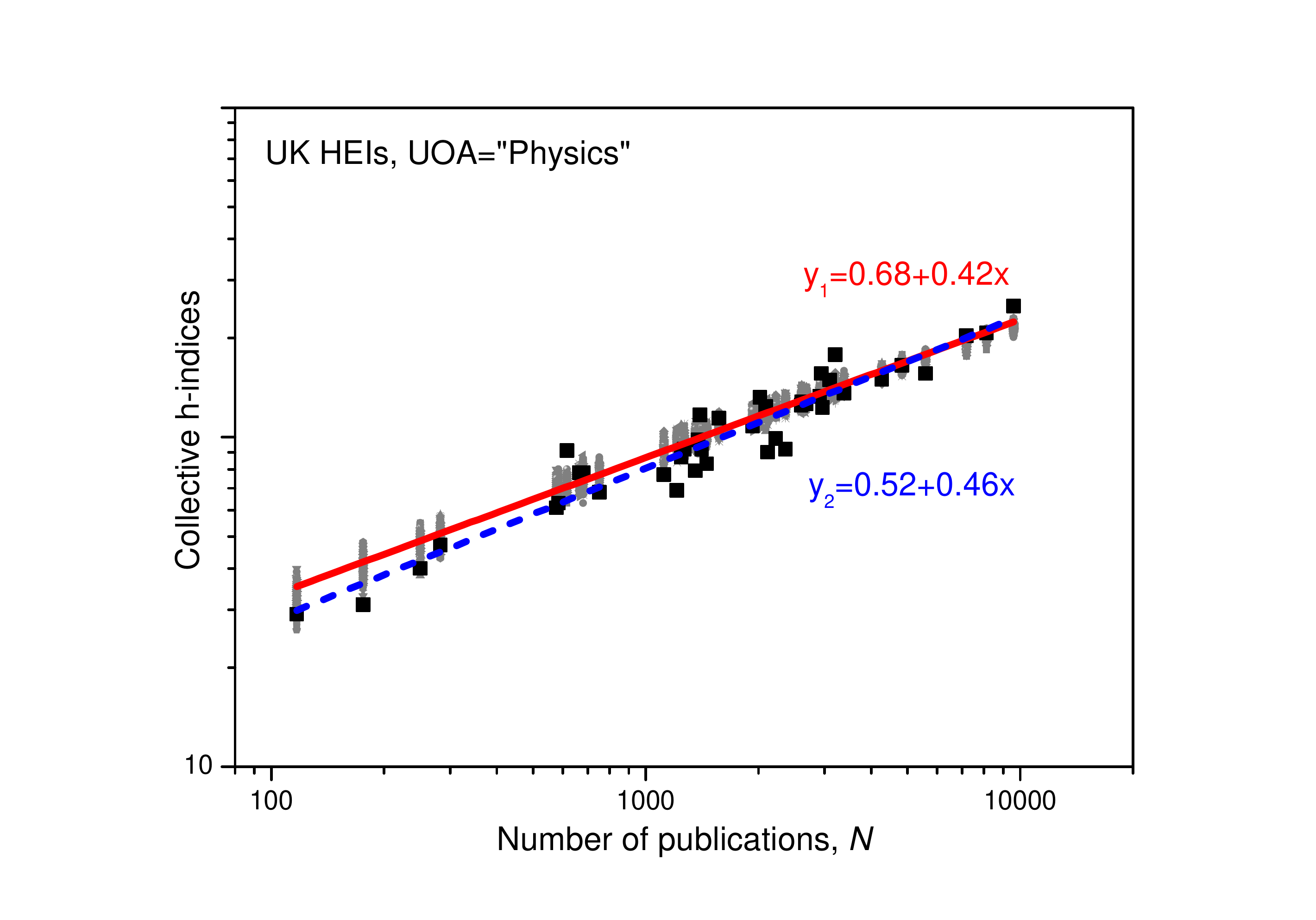}}
	\caption{Values of group \emph{h-indices} vs group sizes for 41 HEIs in {the} UK submitted to RAE2008 (UOA ``Physics''): for the reshuffled data (grey color) and for the
real one (black symbols). Both data tend to follow a power-law dependency, Eq.~(\ref{eq1_beta}). 
 {The solid red line {is a fit to}  all reshuffled data put in one bag implying $\beta \approx 0.42$ while
	the broken blue line is for real data with $\beta\approx 0.46$. Both results are statistically significant for $\alpha=0.01$}.}
	\label{Fig_hreshuffled_UK}
\end{figure}

We again randomly re-allocated outputs to different institutions, taking their sizes into account through the volume of outputs they generate (over 95 thousand papers).
These original data and their randomly reshuffled counterparts are presented in Fig.~\ref{Fig_hreshuffled_UK}. The notation here is the same as in Fig.~\ref{Fig_hreshuffled_MON}.
%
The remarkable similarity between the black and grey data points indicates that the departmental \emph{h-indices} emanating from research outputs generated by selected universities
is {not significantly} different to departmental \emph{h-indices}  that would emanate by  random re-allocation of those outputs.
%
One can notice that actuall data is even closer to reshuffled comparing to the previous case study: {the} Spearman correlation coefficient between real and reshuffled
\emph{h-index} values {is about} $0.94$ on average.
This is {possibly} due to disciplinary homogeneity of publication data.

\begin{figure}[ht]
	\centerline{\includegraphics[width=0.8\textwidth]{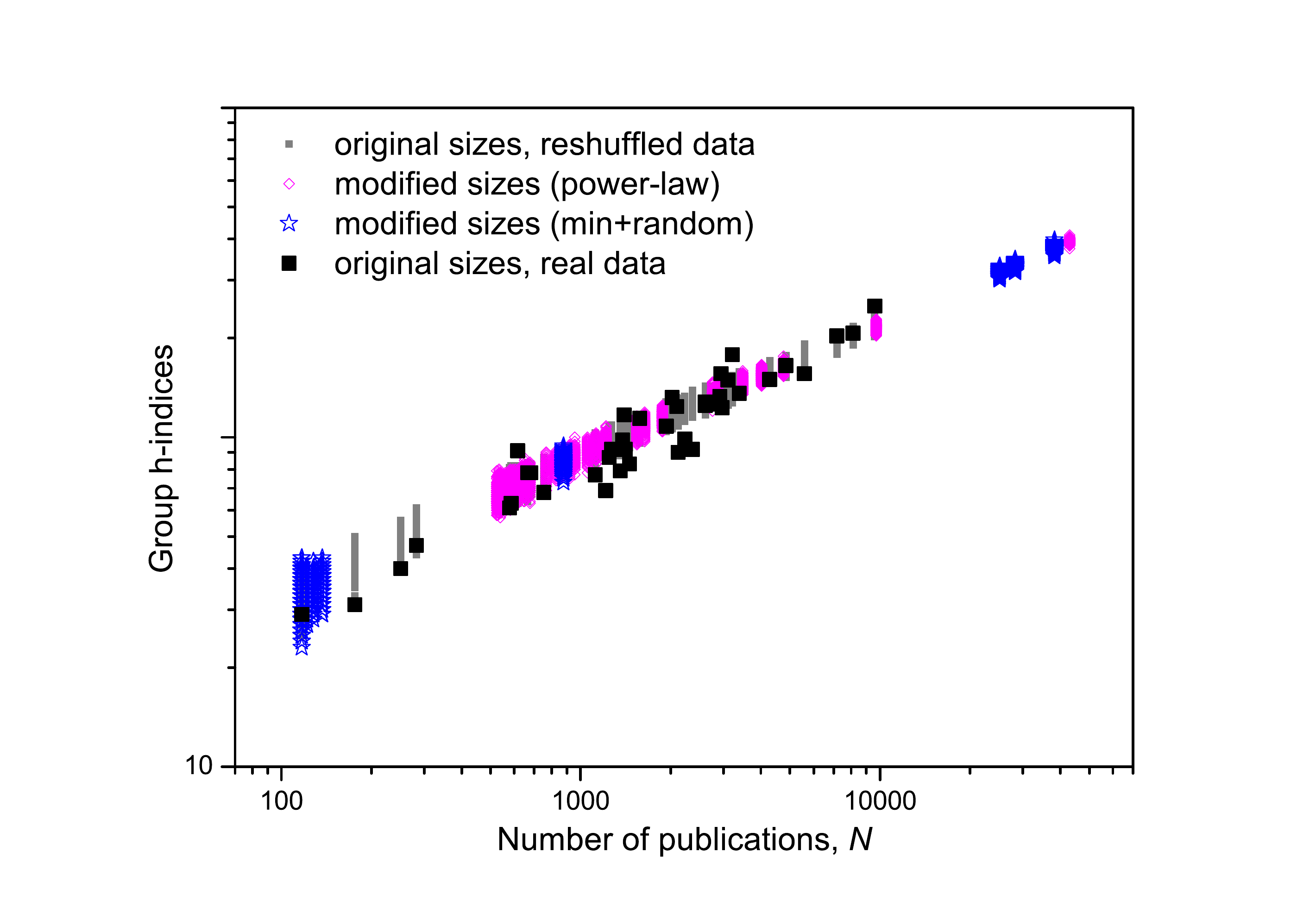}}
	\caption{Values of group \emph{h-indices} vs group sizes: real data for 41 HEIs in {the} UK submitted to RAE2008 (UOA ``Physics'') are shown by black symbols; results for
reshuffled citations using real group sizes are shown by grey squares; reshuffling results for modified distributions of groups sizes: (i) following power-law (magenta diamonds) or
(ii) random with the bottom limit defined (blue stars). }
	\label{Fig_modified_UK}
\end{figure}

To summarize,  Figs.~\ref{Fig_hreshuffled_MON} and \ref{Fig_hreshuffled_UK} {suggest}
{(nonlinear) correlation between group} \emph{h-indices} and number of research outputs.
{Moreover, independently of the way in which empirical data are collected, the same phenomenon occurs and similar values of exponent are found.}
This can be {also} seen in Fig.~\ref{Fig_modified_UK}, where the sizes of research groups are {not taken from real data} but derived from
power-law or random distributions.
The shape of characteristic curve is {also} not dependent on this choice.
One can see that the {manner in which} group \emph{h-index} depends on the number of outputs is {primarily determined} by {an}other input
parameter --  citations distribution. This conclusion is in agreement with theoretical findings {from Refs.}\cite{glanzel2006h,glanzel2019springer}.

\section{Discussion}\label{secX}

The \emph{h-index} is an attractive metric, widely used for research assessments {at} different levels of aggregation by science managers.
The reason for this is that it is easy to calculate the \emph{h-index} for any set of publications:
the algorithm is simple and an interpretation is fast.
However, such haste can be dangerous since it {obfuscates} the principal {peculiarity} of the \emph{h-index}: it is neither an absolute, nor a specific indicator.
The \emph{h-index} depends on the number of sources (publications) by definition and therefore it is dependent on the number of producers (authors) of these sources.
{Because} the \emph{h-index} {is not} normalized per capita {it cannot meaningfully compare groups of different sizes}.
While several theoretical models were designed to make it possible to estimate size-corrected values of {the} \emph{h-index},  {normalisation procedures did not gain traction in
numerous managerial quarters, perhaps because of the loss of simplicity}.
This paper attracts attention to this problem {by} providing {a} clear, {and hopefully convincing,} illustration of {the} size-dependency of departmental (or group)
\emph{h-indices}.
Using an approach widely known in the natural sciences in circumstances where  systems are  investigated to identify principal dependencies,
we reshuffled citation data to redistribute it between research units (institutions, departments).
The correlations which originate from special hiring strategies, better level of infrastructure or invested funds, and other factors, were lost during the reshuffling.
The only difference between research units left was size.
{The reshuffled \emph{h-indices} {were found to be only slightly different} from their original counterparts.}
Our results confirmed that  group \emph{h-index} lend advantage to larger groups simply due to their size because of statistical reasons.


{
In section~\ref{sec2} we presented a toy model to illustrate the intuitively obvious fact that the larger a basket is, the larger the number of balls of certain kind can be picked out of it.
Next, we considered} the set of research publications as a collection of balls of many different colors ({indicated} different citation scores). They were distributed
among the institutions of different sizes represented by the baskets of varying capacity. {Moreover, the papers of some kind are frequent while the others (i.e., highly cited, the most favoured ones) are rare, that follows from the known shapes of real citations distributions, e.g., see \cite{glanzel2019springer}. Putting `balls' from all `baskets' together, reshuffling and redistributing them over the `baskets` again, we lost any correlations between publications and any additional factors remaining solely with the sizes of collections.}

Putting `balls' from all `baskets' together, reshuffling and redistributing them over the `baskets` again, we lost any correlations between publications and any additional factors
remaining solely with the sizes of collections. This allowed us to separate the factor of scale and assess its impact on resulting collective \emph{h-index}. To check this for real
data, we performed two case studies.
{(i)} The collection of papers (balls) with different citation counts (colors) published by Ukrainian HEIs of different sizes (backets) was analysed in
section~\ref{sec4UKR}. {(ii)} Another case study described in section~\ref{sec4UK} based on the collection on publications by the set of research groups from
universities and higher education institutes (HEI) in the UK in the field of Physics (here the limited publication window is considered). In both cases it was demonstrated that
larger HEIs have higher chances to get better scores even for randomly reshuffled data. This indicates that the \emph{h-index} calculated for group output in a straighforward way is
an ineffective measure of group research impact.

If the \emph{h-index} is to be used for comparison as a collective metric, results presented here demonstrate that it should be normalised according to size.
Our approach opens a way to achieve that; the size-dependent averages of the reshuffled group \emph{h-index} scores shown in Figs.~\ref{Fig_hreshuffled_MON} and
\ref{Fig_hreshuffled_UK} present obvious benchmarks against which to compare cores of research groups of given sizes.
In other words, rather than comparing the raw collective \emph{h-index} score of one university to that of another (of different size), one can compare this simple metric to the
average score coming from reshuffled data for groups of commensurate size.
This has the added advantage of environment-specific comparisons --- the benchmark curve for Fig.~\ref{Fig_hreshuffled_MON} differs from that in  Fig.~\ref{Fig_hreshuffled_UK} so
normalisation can be tailored to specific national environments (Ukraine and the UK in this instance) and, desirably, disciplines.
We hope to develop and improve this normalisation approach in follow-on research and present it in a manner accessible to all.

{To conclude, it is natural to expect that larger research units demonstrate better performance due to many factors: more competitive environment and greater chance
to get excellent experts from the larger pool of candidates, better infrastructure and ability to attract more funds, etc. However, it is worth noting that large institutions have
also additional advantage due to sampling bias. Our work provides a means to separate and describe this effect. This is another contribution to the discussion of group size as
crucial factor in measurements of research quality. Moreover, for some practical tasks, e.g., allocating of funds to support the pockets of research excellence, the size of research
units becomes one of dominant factors, as it was shown in our previous works, e.g., see \cite{kenna2011critical,kenna2012}. }
{We believe that our work contributes to the promotion of the contextualized scientometrics that is not targetted on providing the universal a priori list of indicators to be used but on choosing indicators proper for each particular task based on the understanding \cite{Waltman2019}. In this context, group \emph{h-index} is not good or bad per se, but it has to be used and interpreted taking into account its peculiarities, e.g., size-dependency.}

\section*{Acknowledgements}
{The a}uthors thank to Bertrand Berche and Ihor Mryglod for useful discussions. OM thanks to Serhii Nazarovets for inspiration which came from the previous joint work.

\end{document}